\documentclass[11pt,tightenlines,onecolumn,pre,floatfix, superscriptaddress]{revtex4}

\usepackage{url,hyperref,lineno,microtype}

\usepackage{amsmath}				% Math packages (American Mathematical Society)
\usepackage{amssymb}				% "
\usepackage{amsfonts}			% "
\usepackage{mathtools}			% More math
\usepackage{bm}					% Formatting of math symbols

\newcommand{\ti}{{\tau_{\mathrm i}}}
\newcommand{\tf}{{\tau_{\mathrm f}}
\renewcommand{\ss}{{\mathrm{ss}}}}

\newcommand{\T}{\mathsf{T}}
\newcommand{\repl}{\longrightarrow} % replace

\newcommand{\ind}{{\mathrm{ind}}}

\newcommand{\Det}{\operatorname{Det}}
\newcommand{\e}{\mathrm{e}}		% Euler number

\newcommand{\given}{\,|\,}	% conditioning (of probabilities, expectations, ...)

\newcommand{\lmat}{\left( \begin{matrix}}
\newcommand{\rmat}{\end{matrix} \right)}

\newcommand*{\VEC}[1]{\boldsymbol{#1}}

\newcommand*{\kB}{k_{\mathrm{B}}}

\renewcommand{\d}{\mathrm{d}}	% differential
\newcommand{\D}{\mathcal{D}}

\newcommand*{\Da}{D_{\mathrm{a}}}
\newcommand*{\ta}{\tau_{\mathrm{a}}}

\newcommand*{\traj}[1]{\underline{#1}}
\newcommand*{\Traj}[1]{\overline{#1}}
\newcommand*{\pp}[1]{\mathfrak{p}[{#1}]}

\newcommand*{\TrajR}[1]{\overline{\tilde{#1}}}
\newcommand*{\ppR}[1]{\tilde{\mathfrak{p}}[{#1}]}

\newcommand{\I}{\mathrm{i}}			% initial
\newcommand{\F}{\mathrm{f}}			% final

\begin{document}

\author{Lennart Dabelow}
\email{ldabelow@physik.uni-bielefeld.de}
\affiliation{Faculty of Physics, 
Bielefeld University, 
33615 Bielefeld, Germany}

\author{Ralf Eichhorn}
\email{eichhorn@nordita.org}
\affiliation{Nordita, Royal Institute of Technology and Stockholm University, SE-106 91 Stockholm, Sweden}

\title{Irreversibility in active matter:\\ General framework for active Ornstein-Uhlenbeck particles}

\begin{abstract}

Active matter systems are driven out of equilibrium by conversion of energy into directed motion locally on the level of the individual constituents. In the spirit of a minimal description, active
matter is often modeled by so-called active Ornstein-Uhlenbeck particles (AOUPs), an extension of \emph{passive} Brownian motion where activity is represented by an additional fluctuating non-equilibrium ``force’’ with simple statistical properties (Ornstein-Uhlenbeck process). While in \emph{passive} Brownian motion, entropy production along trajectories is well-known to
relate to irreversibility in terms of the log-ratio of probabilities to observe a certain particle trajectory
forward in time in comparison to observing its time-reversed twin trajectory,
the connection between these concepts for active matter is less clear. It is therefore of central importance to provide explicit expressions for the irreversibility of active particle trajectories based on measurable quantities alone, such as the particle positions. In this technical note, we derive a general expression for the irreversibility of AOUPs in terms of path probability ratios (forward versus backward path), extending recent results from
[PRX \textbf{9}, 021009 (2019)] by allowing for arbitrary initial particle distributions and states of the active driving.
\end{abstract}

\maketitle

\section{Introduction}

Irreversible thermodynamic processes are characterized by a positive entropy change in their ``universe'',
i.e., in the combined system of interest and its environment \cite{callen2006thermodynamics}.
In macroscopic (equilibrium) thermodynamics,
where entropy is a state variable,
this change usually refers to the difference between the entropy in the final state of the
``universe'' reached at the end of the process and in the initial state from where it started.
In small mesoscopic systems on the micro- and nanometer scale, such as
a colloidal Brownian particle diffusing in an aqueous solution,
it has been established within the framework of stochastic thermodynamics
\cite{Seifert:2008stp,Jarzynski:2011eai,Seifert:2012stf,van2015ensemble,seifert2018stochastic}
that the total entropy change should be
evaluated from the entropy produced in the system and in its thermal environment
\emph{along the specific trajectory} the system follows during the process.
This procedure remains valid even when the system
is far from equilibrium,
for example due to persistent currents or because it is driven by an
external protocol realizing the thermodynamic process.
The omnipresence of thermal fluctuations on the mesoscopic scale
leads to a distribution of possible paths the system can take to go
from the initial to the final state, and, accordingly to a distribution of entropy changes.
A central result in stochastic thermodynamics is
that the total entropy change $\Delta S$ along a specific
realization of the system path
(divided by Boltzmann's constant $k_{\mathrm B}$) equals
the log-ratio of probabilities for
observing that specific path versus observing
the same path
in a time-reversed manner, i.e., traversing the same trajectory, but
from the final state to the original initial state
\cite{Seifert:2008stp,Jarzynski:2011eai,Seifert:2012stf,van2015ensemble}.
As a direct consequence, the total entropy change $\Delta S$ fulfills a so-called fluctuation theorem,
$\langle \exp (-\Delta S/\kB) \rangle = 1$
(the angular brackets denote an average over all trajectories connecting the initial and final states),
which can be viewed as a generalization of the second law of thermodynamics
to the non-equilibrium realm
when deviations from equilibrium are induced
by \emph{externally applied} forces or gradients.

A fundamentally different class of non-equilibrium systems are so-called ``active particles'',
like Janus colloids with catalytic surfaces or bacteria
\cite{romanczuk2012active,Cates:2012dtw,elgeti2015physics,Bechinger:2016api,patteson2016active},
which have the
ability to locally convert energy into self-propulsion, i.e., they move
independently of external forces or thermal fluctuations.
The source of non-equilibrium
is the energy-to-motion conversion process on the level of the individual particle.
This out-of-equilibrium process produces entropy, but
the various degrees of freedom maintaining the self-propulsion are
usually not observable in typical experiments with
active particles,
such that this entropy production can in general not be quantified.
Moreover, for the (collective) behavior of active particles emerging from self-propulsion,
as described, e.g., in \cite{Tailleur:2008smo,speck2014effective,takatori2014swim,Farage:2015eii},
the details of the propulsion mechanism and the amount of dissipation connected with
it are largely irrelevant. 
In analogy to the stochastic thermodynamics of passive Brownian particles,
a central question in active matter is therefore how the path probabilities
for translational degrees of freedom of the active particles and the associated
log-ratio of forward versus backward path probabilities is connected to irreversibility and
entropy production \cite{Fodor:2016hff,nardini2017entropy,dabelow2019irreversibility}.
We remark that this
is an ongoing debate \cite{Fodor:2016hff,marconi2017heat,mandal2017entropy,
Puglisi:2017crf,caprini2018comment,mandal2018mandal,caprini2019entropy}
which we will not resolve here.
Rather, we will provide a central step towards an understanding of the role of the
path probability ratio in active matter by providing exact analytical expressions for a simple
but highly successful and well-established
\cite{Fily:2012aps,Farage:2015eii,Maggi:2014gee,Argun:2016nbs,maggi2017memory,chaki2018entropy,Marconi:2015tas,shankar2018hidden}
model of active matter,
namely the active Ornstein-Uhlenbeck particle (AOUP)
\cite{Fodor:2016hff,marconi2017heat,mandal2017entropy,Puglisi:2017crf,
koumakis2014directed,szamel2014self,szamel2015glassy,maggi2015multidimensional,Flenner:2016tng,paoluzzi2016critical,Marconi:2016vdi,szamel2017evaluating,sandford2017pressure,caprini2018linear,fodor2018statistical,dabelow2019irreversibility,
dal2019linear,caprini2019entropy,bonilla2019active}.
In this model, self-propulsion is realized via a fluctuating ``driving force''
in the equations of motion \cite{romanczuk2012active,Bechinger:2016api}
with Gaussian distribution and exponential time-correlation
(see Section \ref{sec:model}).
By integrating out these active fluctuations, we
derive an explicit analytical expression for the path weight of an AOUP,
valid for arbitrary values of the model parameters, arbitrary finite duration
of the particle trajectory and arbitrary initial distributions of particle positions and
active fluctuations (see Section \ref{sec:PathWeightArbitrary}).
Using this path weight, we then derive the irreversibility measure in 
form of the log-ratio of forward versus backward path probabilities
\revised{and comment on its physical implications} (Section \ref{sec:irr}).
Before establishing these general results, we briefly recall earlier findings
from \cite{dabelow2019irreversibility} for
independent initial conditions of particle positions and active fluctuations,
see Section \ref{sec:PathWeightIndependent}.
We conclude with a short discussion in Section \ref{sec:discussion},
including potential applications of our results.

\section{Setup}
\label{sec:setup}

\subsection{Model}
\label{sec:model}
The model for an active Ornstein-Uhlenback particle (AOUP)
consists in a standard overdamped Langevin equation for
a passive Brownian particle at position $\VEC{x}$ in $d$ dimensions
with an additional fluctuating force, which represents
the active self-propulsion and which we denote by $\sqrt{2 \Da}\,\VEC{\eta}(t)$,
\begin{equation}
\label{eq:EOM}
	\dot{\VEC{x}}(t) = \frac{1}{\gamma} \VEC{f}(\VEC{x}(t), t) + \sqrt{2 \Da} \, \VEC{\eta}(t) + \sqrt{2 D} \, \VEC{\xi}(t)
\, .
\end{equation}
Here, the dot denotes the time-derivative, $\gamma$ is the viscous friction coefficient,
$\VEC{f}(\VEC{x},t)$ represents externally applied forces 
(conservative or non-conservative, and possibly time-dependent).
Furthermore,
$\VEC{\xi}(t)$ are mutually independent Gaussian white noise sources modeling thermally fluctuating forces with
$\delta$-correlation in time, i.e., $\langle \xi_i(t) \rangle = 0$,
$\langle \xi_i(t)\xi_j(t') \rangle = \delta_{ij}\delta(t-t')$,
and $D$ is the particle diffusion coefficient, related to the temperature
$T$ of the thermal bath via Einstein's relation $D=\kB T/\gamma$.
All bold-face letters represent $d$-dimensional vectors with
components usually labeled by subscripts $i$, $j$, etc.
In analogy to the thermal fluctuations, we denote the strength of the active
fluctuations $\VEC{\eta}(t)$ by $\sqrt{2\Da}$ with an active ``diffusion coefficient'' $\Da$.
For an AOUP, the active fluctuations follow a Gaussian process with exponential
time-correlations, which can be generated by a so-called Ornstein-Uhlenbeck process,
\begin{equation}
\label{eq:LangevinActiveNoise}
	\dot{\VEC{\eta}}(t) = -\frac{1}{\ta} \VEC{\eta}(t) + \frac{1}{\ta} \VEC{\zeta}(t)
\, ,
\end{equation}
where $\ta$ is the correlation time of the active noise fluctuations, i.e.,
\begin{equation}
\langle \eta_i(t)\eta_j(t') \rangle = \frac{\delta_{ij}}{2\ta} e^{-|t-t'|/\ta}
\, .
\end{equation}

\subsection{Central quantity of interest}
Our central goal is to evaluate the path weight $\pp{\traj{\VEC{x}} \given \VEC{x}_\I}$
for particle positions alone,
conditioned on the initial position $\VEC{x}_\I$ for an \emph{arbitrary}
initial distribution $p_\I(\VEC{\eta}_\I \given \VEC{x}_\I)$
of the active fluctuations given the specific value $\VEC{x}_\I$.
By definition, we can write this path weight as
\begin{equation}
\label{eq:PathWeight}
\pp{\traj{\VEC{x}} \given \VEC{x}_\I}
	= \int \D\Traj{\VEC{\eta}} \; \pp{ \traj{\VEC{x}}, \traj{\VEC{\eta}} \given \VEC{x}_\I, \VEC{\eta}_\I } \,
	  p_\I(\VEC{\eta}_\I \given \VEC{x}_\I)
\, ,
\end{equation}
where the path integral over $\Traj{\VEC{\eta}}:=\{\VEC{\eta}(t)\}_{t=\ti}^\tf$
includes the initial configuration $\VEC{\eta}_\I$,
whereas the notation $\traj{\VEC{\eta}} := \{\VEC{\eta}(t)\}_{t>\ti}^\tf$ denotes the
same history of active fluctuations \emph{without} the initial configuration $\VEC\eta_\I$,
and similarly for $\traj{\VEC x} := \{\VEC{x}(t)\}_{t>\ti}^\tf$.
Moreover,
\begin{equation}
\label{eq:PathWeightJoint}
\pp{ \traj{\VEC{x}}, \traj{\VEC{\eta}} \given \VEC{x}_\I, \VEC{\eta}_\I }
	\propto \exp \left\{ -\int_\ti^\tf \!\! \d t
		\left[ \frac{(\dot{\VEC{x}}_t - \VEC{v}_t - \sqrt{2\Da} \VEC{\eta}_t )^2}{4D}
			 + \frac{(\ta \dot{\VEC{\eta}}_t + \VEC{\eta}_t)^2}{2}
			 + \frac{\VEC{\nabla} \cdot \VEC{v}_t}{2}
		\right] \right\}
\end{equation}
is the standard Onsager-Machlup path weight
\cite{Onsager:1953fai,Machlup:1953fai,cugliandolo2017rules}
for the joint process $(\traj{\VEC{x}}, \traj{\VEC{\eta}})$,
where we use the shorthand notation 
$\VEC{v}_t = \VEC{f}_t/\gamma = \VEC{f}(\VEC{x}(t), t)/\gamma$ and
$\VEC{x}_t \equiv \VEC{x}(t)$, $\VEC{\eta}_t \equiv \VEC{\eta}(t)$, etc.
The technical challenge consists in performing the integral over the active fluctuations
$\Traj{\VEC{\eta}}$ without explicitly specifying the initial distribution
$p_\I(\VEC{\eta}_\I \given \VEC{x}_\I)$.

\section{Path weight for independent initial conditions}
\label{sec:PathWeightIndependent}

\subsection{The results from \cite{dabelow2019irreversibility}}
\label{sec:resultsPRX}
We start by summarizing the main results from \cite{dabelow2019irreversibility}.
In~\cite{dabelow2019irreversibility} we gave the path weight
for trajectories $\Traj{\VEC{x}} = \{ \VEC{x}(t) \}_{t=0}^\tau$,
running from initial time $\ti = 0$ to final time $\tf = \tau$,
assuming that the active noise is initially independent of the
particle positions and in its steady state,
i.e., $p_\I(\VEC{\eta}_0 \given \VEC{x}_0) = p_{\mathrm{ss}}(\VEC{\eta}_0) = \sqrt{\ta/\pi} \, \e^{-\ta \VEC{\eta}_0^2}$.
We found
\begin{align}
\label{eq:PathWeightInd:0Tau}
\mathfrak{p}^{\mathrm{ind}}_{(0, \tau]}[\traj{\VEC{x}} \given \VEC{x}_0]
& \propto \exp
	\left\{ -\frac{1}{4D} \int_0^\tau \!\! \d t \int_0^\tau \!\! \d t' \,
		\left[ \dot{\VEC{x}}_t - \VEC{v}_t \right]^\T
		\left[ \delta(t - t') - \tfrac{\Da}{D} \Gamma^{\mathrm{ind}}_{[0,\tau]}(t, t') \right]
		\left[ \dot{\VEC{x}}_{t'} - \VEC{v}_{t'} \right] 
	\right.
\nonumber \\
& \qquad\qquad
	\left. \mbox{}
		- \frac{1}{2} \int_0^\tau \!\! \d t \; \VEC{\nabla} \cdot \VEC{v}_t
	\right\}
\, ,
\end{align} 
with the memory kernel
\begin{equation}
\label{eq:GammaInd:0Tau}
\Gamma^\ind_{[0, \tau]}(t, t') :=
	\left( \frac{1}{2 \ta^2 \lambda} \right)
	\frac{ \kappa_+^2 \e^{-\lambda |t - t'|} 
			+ \kappa_-^2 \e^{-\lambda(2\tau - |t - t'|)}
			- \kappa_+ \kappa_- \left[ \e^{-\lambda(t+t')} + \e^{-\lambda(2\tau - t - t')} \right] }
	     { \kappa_+^2 - \kappa_-^2 \e^{-2 \lambda\tau} },
\end{equation}
where $\lambda := \sqrt{1 + \Da/D} / \ta$ and $\kappa_{\pm} := 1 \pm \sqrt{1 + \Da / D}$.

\subsection{Stationary-state scenario}
\label{sec:PathWeightIndependentSS}
If we have a trajectory $\Traj{\VEC{x}} = \{ \VEC{x}(t) \}_{t=\ti}^{\tf}$
running from arbitrary times $\ti$ to $\tf$ instead,
we can shift time as $t \repl t - \ti$ and identify
$\tau = \tf - \ti$ as the duration of the trajectory to convert
$(0,\tau]$ path weights to those running from $\ti$ to $\tf$.
Performing these replacements, the memory kernel~\eqref{eq:GammaInd:0Tau} turns into
\begin{equation}
\label{eq:GammaInd}
\Gamma^\ind_{[\ti, \tf]}(t, t') :=
	\left( \frac{1}{2 \ta^2 \lambda} \right)
	\frac{ \kappa_+^2 \e^{-\lambda |t - t'|}
			+ \kappa_-^2 \e^{-\lambda[2(\tf - \ti) - |t - t'|]}
			- \kappa_+ \kappa_- \left[ \e^{-\lambda(t+t' - 2 \ti)}
			+ \e^{-\lambda(2\tf - t - t')} \right] }
		 { \kappa_+^2 - \kappa_-^2 \e^{-2 \lambda(\tf - \ti)} }
\, .
\end{equation}
Consequently, the corresponding path weight for a trajectory starting at $\VEC{x}_\I$ at time $\ti$ reads
\begin{align}
\label{eq:PathWeightInd}
\mathfrak{p}^\ind_{(\ti, \tf]}[\traj{\VEC{x}} \given \VEC{x}_\I]
& \propto \exp
	\left\{ -\frac{1}{4D} \int_\ti^\tf \!\! \d t \int_\ti^\tf \!\! \d t' \,
		\left[ \dot{\VEC{x}}_t - \VEC{v}_t \right]^\T
		\left[ \delta(t - t') - \tfrac{\Da}{D} \Gamma^\ind_{[\ti,\tf]}(t, t') \right]
		\left[ \dot{\VEC{x}}_{t'} - \VEC{v}_{t'} \right] 
	\right. \notag
\\ & \qquad\qquad
	\left. \mbox{}
		- \frac{1}{2} \int_\ti^\tf \!\! \d t \; \VEC{\nabla} \cdot \VEC{v}_t
	\right\}
\, .
\end{align} 
Letting $\ti \to -\infty$ (stationary-state scenario), the memory kernel becomes
\begin{equation}
\label{eq:GammaIndep:SS}
	\Gamma^\ind_{(-\infty, \tf]}(t, t') = \frac{1}{2 \ta^2 \lambda} \left[ \e^{-\lambda \lvert t - t' \rvert} - \frac{\kappa_-}{\kappa_+} \e^{-\lambda( 2\tf - t - t' )} \right]
\, .
\end{equation}
For ``infinitely long'' stationary-state trajectories, for which also $\tf \to \infty$, this expression further reduces to
\begin{equation}
\label{eq:GammaIndep:SSinf}
	\Gamma^\ind_{(-\infty, \infty)}(t, t') = \frac{1}{2 \ta^2 \lambda} \e^{-\lambda \lvert t - t' \rvert}
\, .
\end{equation}
The latter special case has been derived independently in \cite{caprini2019entropy} via Fourier transformation,
see eq.~(25) in \cite{caprini2019entropy}, in order to analyze ``entropy production'' based on
path-probability ratios.
Similar Fourier-transform techniques for Langevin systems have been used in \cite{zamponi2005fluctuation}
for deriving a fluctuation relation at large times, with findings
for the non-local ``inverse temperature'' as integration kernel in
the ``entropy production'' corresponding
to those in \cite{caprini2019entropy}, and to our \eqref{eq:GammaIndep:SSinf}.

\section{Path weight for arbitrary initial conditions}
\label{sec:PathWeightArbitrary}

In this section, we generalize the path weight~\eqref{eq:PathWeightInd}
to allow for arbitrary joint initial distributions $p_\I(\VEC{x}_\I, \VEC{\eta}_\I)$ 
of particle positions and active fluctuations.
Keeping in mind that we can time-shift final results between trajectories running during a
time interval $(0,\tau]$ and during arbitrary intervals $(\ti,\tf]$ as in Sec.~\ref{sec:PathWeightIndependentSS},
we here consider without loss of generality  
trajectories with $\ti = 0$ and $\tf = \tau$.
For notational simplicity we drop the subscripts $(0,\tau]$ or $[0,\tau]$ on $\mathfrak{p}$ and $\Gamma$.

We start in Sec.~\ref{sec:p0etaGaussian}
by first calculating $\Gamma$
for a general Gaussian initial distribution of $\VEC{\eta}_0$ independent of $\VEC{x}_0$,
which has variance $\sigma^2$ and is centered at $\hat{\VEC{\eta}}_0$,
\begin{equation}
\label{eq:pGh}
p_{\hat{\VEC{\eta}}_0,\sigma}(\VEC{\eta}_0)
= \frac{1}{\sqrt{2 \pi \sigma^2}} \e^{-(\VEC{\eta}_0 - \hat{\VEC{\eta}}_0)^2/2\sigma^2}
\, .
\end{equation}
Then, in Sec.~\ref{sec:p0etaArbitrary},
we show how this result can be used to cover any arbitrary initial
distribution $p_\I(\VEC{\eta}_0 \given \VEC{x}_0)$.

\subsection{Gaussian initial distribution}
\label{sec:p0etaGaussian}
With \eqref{eq:PathWeight}, \eqref{eq:PathWeightJoint}, and the initial distribution
$p_\I(\VEC{\eta}_\I \given \VEC{x}_\I) = p_\I(\VEC{\eta}_0 \given \VEC{x}_0) = p_\I(\VEC{\eta}_0) = p_{\hat{\VEC{\eta}}_0,\sigma}(\VEC{\eta}_0)$ 
from \eqref{eq:pGh}, the
path weight we want to evaluate reads
\begin{align}
\label{eq:PathWeighth0sigma}
\mathfrak{p}^\ind_{\hat{\VEC{\eta}}_0,\sigma}[\traj{\VEC{x}} \given \VEC{x}_0]
	& \propto \frac{1}{\sqrt{2 \pi \sigma^2}}  \int \D\Traj{\VEC{\eta}} \;
	\exp \left\{ -\int_0^\tau \!\! \d t
		\left[ \frac{(\dot{\VEC{x}}_t - \VEC{v}_t - \sqrt{2\Da} \VEC{\eta}_t )^2}{4D}
			 + \frac{(\ta \dot{\VEC{\eta}}_t + \VEC{\eta}_t)^2}{2}
			 + \frac{\VEC{\nabla} \cdot \VEC{v}_t}{2}
		\right] \right.
\nonumber \\
& \qquad\qquad\qquad\qquad\qquad\qquad \left. \mbox{}
			 - \frac{(\VEC{\eta}_0 - \hat{\VEC{\eta}}_0)^2}{2\sigma^2}
		\right\}
\, .
\end{align}
The superscript $\ind$ emphasizes again that we use statistically independent initial conditions for $\VEC{x}_0$ and $\VEC{\eta}_0$.
After partial integration of the $\dot{\VEC{\eta}}_t$ terms,
similarly as in~\cite{dabelow2019irreversibility},
we can express the path integral as
\begin{align}
\label{eq:PathWeightGauss:BeforeEtaInt}
\mathfrak{p}^\ind_{\hat{\VEC{\eta}}_0,\sigma}[ \traj{\VEC{x}} \given \VEC{x}_0 ]
	& \propto \frac{1}{\sqrt{2\pi\sigma^2}}
	\exp\left\{ -\int_0^\tau \!\! \d t
		\left[ \frac{(\dot{\VEC{x}}_t - \VEC{v}_t)^2}{4D} + \frac{\VEC{\nabla} \cdot \VEC{v}_t}{2} \right]
		- \frac{ \hat{\VEC{\eta}}_0^2 }{2\sigma^2}
	\right\}
\nonumber \\
& \hspace*{-5ex}\mbox{}\times
	\int \D\Traj{\VEC{\eta}} \;
	\exp\left\{
		- \frac{1}{2} \int_0^\tau \!\! \d t \int_0^\tau \!\! \d t' \; \VEC{\eta}_t^\T V_\sigma(t, t') \VEC{\eta}_{t'}
		+ \int_0^\tau \!\! \d t \; \VEC{\eta}_t^\T \left[ \frac{\sqrt{2\Da}}{2D}
				(\dot{\VEC{x}}_t - \VEC{v}_t) + \delta(t) \frac{\hat{\VEC{\eta}}_0}{\sigma^2} \right]
	\right\}
\, ,
\end{align}
with the differential operator
\begin{equation}
\label{eq:PathWeightGauss:V}
V_\sigma(t, t')
	:= \delta(t - t') \left[
		- \ta^2 \partial_{t'}^2 + 1 + \frac{\Da}{D}
		+ \delta(t') \left( -\ta^2 \partial_{t'} - \ta + \frac{1}{\sigma^2} \right)
		+ \delta(\tau - t') \left(\ta^2 \partial_{t'} + \ta \right)
	\right]
\, .
\end{equation}
Performing the Gaussian integral over $\Traj{\VEC{\eta}}$ in~\eqref{eq:PathWeightGauss:BeforeEtaInt}, we obtain
\begin{align}
\label{eq:PathWeightGauss:AfterEtaInt}
\mathfrak{p}^\ind_{\hat{\VEC{\eta}}_0,\sigma}[ \traj{\VEC{x}} \given \VEC{x}_0 ]
	& \propto \frac{(\Det V_\sigma)^{-1/2}}{\sqrt{2\pi\sigma^2}}
	\exp\left\{ -\frac{1}{4D} \int_0^\tau \!\! \d t \int_0^\tau \!\! \d t' \;
		\left( \dot{\VEC{x}}_t - \VEC{v}_t \right)^\T
		\left[ \delta(t - t') - \tfrac{\Da}{D} \Gamma_\sigma(t, t') \right]
		\left( \dot{\VEC{x}}_{t'} - \VEC{v}_{t'} \right) 
	\right.
\nonumber \\
	& \qquad\qquad\qquad\qquad\qquad
	\left. \mbox{} %\vphantom{\int_0^\tau  \frac{\sqrt{2 \Da}}{2 D \sigma^2}}
		+ \int_0^\tau \!\! \d t \left[
			\frac{\sqrt{2\Da}}{2D} \left( \dot{\VEC{x}}_t - \VEC{v}_t \right)^\T
			\frac{ \Gamma_\sigma(t, 0) }{ \sigma^2 } \hat{\VEC{\eta}}_0 - \frac{\VEC{\nabla} \cdot \VEC{v}_t}{2}
		\right]
	\right.
\nonumber \\
	& \qquad\qquad\qquad\qquad\qquad
	\left. \mbox{} %\vphantom{\int_0^\tau \frac{\sqrt{2 \Da}}{2 D \sigma^2}}
		+ \left[ \frac{\Gamma_\sigma(0, 0)}{\sigma^2} - 1 \right] \frac{\hat{\VEC{\eta}}_0^2}{2\sigma^2}
	\right\}
\, ,
\end{align}
where $\Gamma_\sigma(t, t')$ denotes the operator inverse of $V_\sigma(t, t')$
in the sense that
$\int_0^\tau \d t' \; V_\sigma(t,t') \Gamma_\sigma(t',t'') = \delta(t - t'')$.
It can be constructed similarly to the procedure in \cite{dabelow2019irreversibility}.
In particular, we can also write $\Gamma_\sigma(t, t^\prime) = G(t,t') + H_\sigma(t,t')$.
Here $G(t,t')$ is the Green's function defined by
$[-\tau_a^2 \partial_t^2 + (1 + \Da/D) ] G(t,t') = \delta(t-t')$
and $G(0, t') = G(\tau, t') = 0$.
The second ingredient, $H(t,t')$, is a solution of the associated homogeneous problem,
$[-\tau_a^2 \partial_t^2 + (1 + \Da/D) ] H(t,t') = 0$,
fixing the boundary terms as prescribed by \eqref{eq:PathWeightGauss:V}.
More details are given the Appendix.
We find
\begin{align}
\label{eq:Gamma:Sigma}
\Gamma_\sigma(t, t')
& = \left( \frac{1}{2 \ta^2 \lambda} \right)
		\left[ \kappa_+ (1 - \sigma^2 \ta \kappa_-) - \kappa_- (1 - \sigma^2 \ta \kappa_+) \e^{-2 \lambda\tau} \right]^{-1}
\nonumber \\
& \qquad \mbox{} \times
	\left[ \kappa_+ (1 - \sigma^2 \ta \kappa_-) \e^{-\lambda \lvert t - t' \rvert}
		+ \kappa_- (1 - \sigma^2 \ta \kappa_+) \e^{-\lambda(2\tau - \lvert t - t'\vert)} 
	\right.
\nonumber \\ 
& \qquad\qquad
	\left. \mbox{}
		- \kappa_+ (1 - \sigma^2 \ta \kappa_+) \e^{-\lambda(t+t')}
		- \kappa_- (1 - \sigma^2 \ta \kappa_-) \e^{-\lambda(2\tau - t - t')}
	\right]
\, .
\end{align}

We note that \eqref{eq:pGh} includes the steady-state distribution,
$p_{\mathrm{ss}}(\VEC{\eta}_0) = \sqrt{\ta/\pi} \, \e^{-\ta \VEC{\eta}_0^2}$,
which arises for the active noise when evolving independently of the Brownian particle,
as a special case for $\hat{\VEC{\eta}}_0 = 0$ and $\sigma^2 = 1/(2\ta)$.
Accordingly, we recover \eqref{eq:PathWeightInd:0Tau} and \eqref{eq:GammaInd:0Tau} when plugging
$\hat{\VEC{\eta}}_0 = 0$ and $\sigma^2 = 1/(2\ta)$ into
\eqref{eq:PathWeightGauss:AfterEtaInt} and \eqref{eq:Gamma:Sigma},
using $1-\kappa_\pm/2 = \kappa_\mp/2$.

\subsection{Arbitrary initial distribution}
\label{sec:p0etaArbitrary}
To cover arbitrary initial distributions
$p_\I(\VEC{\eta}_\I \given \VEC{x}_\I) = p_\I(\VEC{\eta}_0 \given \VEC{x}_0)$
in $\VEC{\eta}_0$, we
introduce a $\delta$-distribution of the form
$\delta(\VEC{\eta}_0 - \hat{\VEC{\eta}}_0) = \lim_{\sigma\to 0} \e^{-(\VEC{\eta}_0 - \hat{\VEC{\eta}}_0)^2/2\sigma^2} / \sqrt{2 \pi \sigma^2}$
and rewrite \eqref{eq:PathWeight} (with $\ti=0$, $\tf=\tau$) as
\begin{align}
\label{eq:PathWeight:SplitOffInit}
\mathfrak{p}[ \traj{\VEC{x}} \given \VEC{x}_0 ]
& = \int \D\Traj{\VEC{\eta}} \; \pp{ \traj{\VEC{x}}, \traj{\VEC{\eta}} \given \VEC{x}_0, \VEC{\eta}_0 } \,
	p_\I(\VEC{\eta}_0 \given \VEC{x}_0)
\nonumber \\
& = \int \D\Traj{\VEC{\eta}} \; \pp{ \traj{\VEC{x}}, \traj{\VEC{\eta}} \given \VEC{x}_0, \VEC{\eta}_0 } \,
	\int \d\hat{\VEC{\eta}}_0 \, \delta(\VEC{\eta}_0 - \hat{\VEC{\eta}}_0) \, p_\I(\hat{\VEC{\eta}}_0 \given \VEC{x}_0)
\nonumber \\
& = \lim_{\sigma\to 0} \int \d\hat{\VEC{\eta}}_0 \; p_\I(\hat{\VEC{\eta}}_0 \given \VEC{x}_0)
	\left[
		\frac{1}{\sqrt{2\pi\sigma^2}} \int \D\Traj{\VEC{\eta}} \; \pp{\traj{\VEC{x}}, \traj{\VEC{\eta}} \given \VEC{x}_0, \VEC{\eta}_0}
		\, \e^{-(\VEC{\eta}_0 - \hat{\VEC{\eta}}_0)^2 / 2 \sigma^2 }
	\right]
\, .
\end{align}
In view of \eqref{eq:PathWeightJoint} we see that the term in brackets is
exactly 
$\mathfrak{p}^\ind_{\hat{\VEC{\eta}}_0,\sigma}[\traj{\VEC{x}} \given \VEC{x}_0]$
as defined in \eqref{eq:PathWeighth0sigma}.
Since we can also write
$\mathfrak{p}[ \traj{\VEC{x}} \given \VEC{x}_0 ]
= \int \d\hat{\VEC{\eta}}_0 \; \pp{ \traj{\VEC{x}},\hat{\VEC{\eta}}_0 \given \VEC{x}_0 }
= \int \d\hat{\VEC{\eta}}_0 \; p_\I(\hat{\VEC{\eta}}_0 \given \VEC{x}_0) \, \pp{ \traj{\VEC{x}} \given \VEC{x}_0,\hat{\VEC{\eta}}_0 }$
we conclude that
\begin{equation}
\label{eq:PathWeightArbAsLimit}
\pp{ \traj{\VEC{x}} \given \VEC{x}_0,\hat{\VEC{\eta}}_0 }
= \lim_{\sigma\to 0} \mathfrak{p}^\ind_{\hat{\VEC{\eta}}_0,\sigma}[\traj{\VEC{x}} \given \VEC{x}_0]
\end{equation}
is the path weight conditioned on an initial position $\VEC{x}_0$
and initial state of the active noise $\hat{\VEC{\eta}}_0$ with arbitrary
distributions.

With the explicit result \eqref{eq:PathWeightGauss:AfterEtaInt} for
$\mathfrak{p}^\ind_{\hat{\VEC{\eta}}_0,\sigma}[\traj{\VEC{x}} \given \VEC{x}_0]$ we thus see
that we have to calculate the $\sigma \to 0$ limit of the expressions
$(\sigma^2 \Det V_\sigma)^{1/2}$, $[\Gamma_\sigma(t,0)/\sigma^2-1]/\sigma^2$, $\Gamma_\sigma(0,0)/\sigma^2$
and $\Gamma_\sigma(t,t')$.
From \eqref{eq:PathWeightGauss:V} we observe that $\sigma^2 V_\sigma$ has a constant
term (independent of $\sigma$) and contributions quadratic in $\sigma$ such that
$(\sigma^2 \Det V_\sigma)^{1/2}$ reduces to an (irrelevant) constant as $\sigma \to 0$.
Next, setting $t'=0$ in \eqref{eq:Gamma:Sigma} and using
$\ta\kappa_+ - \ta\kappa_- = 2\ta^2\lambda$ we get
\begin{equation}
\label{eq:GammaArb:Sigma:t0}
\frac{ \Gamma_\sigma(t, 0) }{ \sigma^2 }
	= \frac{ \kappa_+ \e^{-\lambda t} - \kappa_- \e^{-\lambda(2\tau - t)} }
		   { \kappa_+ (1 - \sigma^2 \ta \kappa_-) - \kappa_- (1 - \sigma^2 \ta \kappa_+) \e^{-2 \lambda\tau} }
	\stackrel{\sigma \to 0}{\longrightarrow}
	  \frac{ \kappa_+ \e^{-\lambda t} - \kappa_- \e^{-\lambda(2\tau - t)} }{ \kappa_+ - \kappa_- \e^{-2 \lambda\tau} }
\, .
\end{equation}
If $t = 0$, too, we obtain
\begin{equation}
\frac{ \Gamma_\sigma(0, 0) }{ \sigma^2 }
	= \frac{ \kappa_+ - \kappa_- \e^{-2 \lambda\tau} }
		   { \kappa_+ (1 - \sigma^2 \ta \kappa_-) - \kappa_- (1 - \sigma^2 \ta \kappa_+) \e^{-2 \lambda\tau} }
\, ,
\end{equation}
such that
\begin{equation}
\label{eq:GammaArb:Sigma:00}
\left[ \frac{ \Gamma_\sigma(0, 0) }{ \sigma^2 } - 1 \right]\frac{1}{\sigma^2}
= \frac{ \kappa_+ \kappa_- \ta (1 - \e^{-2\lambda\tau}) }
	   { \kappa_+ (1 - \sigma^2 \ta \kappa_-) - \kappa_- (1 - \sigma^2 \ta \kappa_+) \e^{-2\lambda\tau} }
\stackrel{\sigma \to 0}{\longrightarrow}
	\frac{ \kappa_+ \kappa_- \ta (1 - \e^{-2\lambda\tau}) }{ \kappa_+ - \kappa_- \e^{-2 \lambda \tau} }
\, .
\end{equation}
Furthermore, we define 
\begin{equation}
\label{eq:GammaArb:0Tau}
\Gamma(t, t') := \lim_{\sigma \to 0} \Gamma_\sigma(t, t')
	= \left( \frac{1}{2\ta^2\lambda} \right)
	  \frac{ \kappa_+ \e^{-\lambda \lvert t - t' \rvert}
	  		+ \kappa_- \e^{-\lambda(2\tau - \lvert t - t'\vert )}
			- \kappa_+ \e^{-\lambda(t+t')}
			- \kappa_- \e^{-\lambda(2\tau - t - t')} }
		   { \kappa_+ - \kappa_- \e^{-2 \lambda\tau} }
\, 
\end{equation}
as the memory kernel for the path weight
conditioned on an arbitrary initial configuration $(\VEC x_0, \hat{\VEC\eta}_0)$
of particle positions and active fluctuations.
Altogether, eq.~\eqref{eq:PathWeightArbAsLimit} for this path weight then becomes
\begin{align}
\label{eq:PathWeightArb:0Tau}
\mathfrak{p}[ \traj{\VEC{x}} \given \VEC{x}_0, \hat{\VEC{\eta}}_0 ]
& \propto \exp\left\{
	-\frac{1}{4D} \int_0^\tau \!\! \d t \int_0^\tau \!\! \d t' \;
		\left(\dot{\VEC{x}}_t - \VEC{v}_t\right)^\T
		\left[ \delta(t - t') - \tfrac{\Da}{D} \Gamma(t, t') \right]
		\left( \dot{\VEC{x}}_{t'} - \VEC{v}_{t'} \right)
	\right.
\nonumber \\
& \qquad\qquad \left. \mbox{}
	+ \int_0^\tau \!\! \d t
		\left[
			\frac{\sqrt{2\Da}}{2D}
			\left( \dot{\VEC{x}}_t - \VEC{v}_t \right)^\T
			\frac{ \kappa_+ \e^{-\lambda t} - \kappa_- \e^{-\lambda(2\tau - t)} }
			     { \kappa_+ - \kappa_- \e^{-2 \lambda\tau} }
			\hat{\VEC{\eta}}_0
			- \frac{\VEC{\nabla} \cdot \VEC{v}_t}{2}
		\right] 
	\right.
\nonumber \\
& \qquad\qquad \left. \mbox{}	
	- \frac{\Da}{2D} \left[
			\frac{ \ta (1 - \e^{-2\lambda\tau}) }{ \kappa_+ - \kappa_- \e^{-2 \lambda \tau} }
		\right] \hat{\VEC{\eta}}_0^2
	\right\} 
\, ,
\end{align}
where we have used $\kappa_+ \kappa_- = -\Da/D$ in the third line.

Finally, we can shift trajectories similarly as in Sec.~\ref{sec:PathWeightIndependentSS}
to obtain the path weight for arbitrary trajectories $\Traj{\VEC{x}} = \{\VEC{x}(t)\}_{t = \ti}^\tf$
conditioned on the joint initial state $(\VEC{x}_\I, \VEC{\eta}_\I)$ of position and active noise,
\begin{align}
\label{eq:PathWeightArb}
\mathfrak{p}_{(\ti, \tf]}[ \traj{\VEC{x}} \given \VEC{x}_\I, \VEC{\eta}_\I ]
	& \propto \exp\left\{ -\frac{1}{4D} \int_\ti^\tf \!\! \d t \int_\ti^\tf \!\! \d t' \;
		\left[ \dot{\VEC{x}}_t - \VEC{v}_t \right]^\T
		\left[ \delta(t - t') - \tfrac{\Da}{D} \Gamma_{[\ti,\tf]}(t, t') \right]
		\left[ \dot{\VEC{x}}_{t'} - \VEC{v}_{t'} \right] 
	\right.
\nonumber \\
    & \qquad\qquad \left. \mbox{}
	+ \int_\ti^\tf \!\! \d t
		\left[
			\frac{\sqrt{2\Da}}{2D}
			\left[ \dot{\VEC{x}}_t - \VEC{v}_t \right]^\T
			\frac{ \kappa_+ \e^{-\lambda (t - \ti)} - \kappa_- \e^{-\lambda(2\tf - t - \ti)} }
				 { \kappa_+ - \kappa_- \e^{-2 \lambda(\tf-\ti)} }
		    \VEC{\eta}_\I
		    - \frac{\VEC{\nabla} \cdot \VEC{v}_t}{2}
		\right] 
	\right.
\nonumber \\
& \qquad\qquad \left. \mbox{}	
	-  \frac{\Da}{2D} \left[
			\frac{ \ta (1 - \e^{-2\lambda(\tf-\ti)}) }{ \kappa_+ - \kappa_- \e^{-2 \lambda(\tf-\ti)} }
		\right] \VEC{\eta}_\I^2
	\right\} 
\end{align}
with
\begin{equation}
\label{eq:GammaArb}
\Gamma_{[\ti, \tf]}(t, t')
	:= \left( \frac{1}{2 \ta^2 \lambda} \right)
	   \frac{ \kappa_+ \e^{-\lambda \lvert t - t' \rvert}
	    	+ \kappa_- \e^{-\lambda(2(\tf - \ti) - \lvert t - t'\vert )} 
		 	- \kappa_+ \e^{-\lambda(t+t' - 2 \ti)} - \kappa_- \e^{-\lambda(2\tf - t - t')} }
			{ \kappa_+ - \kappa_- \e^{-2 \lambda (\tf - \ti)} }
\, .
\end{equation}
Given an initial distribution $p_\I(\VEC{\eta}_\I \given \VEC{x}_\I)$ of the active fluctuations
conditioned on the initial particle position, we can then compute the position-only
path weight of an arbitrary trajectory by averaging over $p_\I(\VEC{\eta}_\I \given \VEC{x}_\I)$,
\begin{equation}
\label{eq:PathWeightArb:PosOnly}
\mathfrak{p}_{(\ti, \tf]}[ \traj{\VEC{x}} \given \VEC{x}_\I ]
	= \int \d\VEC{\eta}_\I \;
		\mathfrak{p}_{(\ti, \tf]}[ \traj{\VEC{x}} \given \VEC{x}_\I, \VEC{\eta}_\I ] \,
		p_\I(\VEC{\eta}_\I \given \VEC{x}_\I)
\, .
\end{equation}
Equations \eqref{eq:PathWeightArb}, \eqref{eq:GammaArb}, \eqref{eq:PathWeightArb:PosOnly} represent 
the first central
result of the present contribution, a general expression for the path weight of active Ornstein-Uhlenbeck particles
in position space only,
for arbitrary trajectories with arbitrary initial and final times and
arbitrary initial distributions. There is no approximation involved, so that our results are valid for any values of
thermal and active noise parameters $D$ and $\Da$, $\ta$.

We expect that the specific initial configuration becomes irrelevant for steady-state trajectories,
i.e., in the limit $\ti \to -\infty$.
As $\ti \to -\infty$, the second line vanishes
in \eqref{eq:PathWeightArb}, because
$\frac{\kappa_+ \e^{-\lambda (t - \ti)} - \kappa_- \e^{-\lambda(2\tf - t - \ti)}}{\kappa_+ - \kappa_- \e^{-2 \lambda(\tf-\ti)}} \to 0$.
The third line enters into the integral over the initial configuration $\VEC{\eta}_\I$
(see \eqref{eq:PathWeightArb:PosOnly})
and thus decouples from the trajectory $\Traj{\VEC{x}}$ resulting in an irrelevant
prefactor.
The only relevant contribution as $\ti \to -\infty$ is therefore the first line in \eqref{eq:PathWeightArb}
with the integral kernel $\Gamma_{[\ti, \tf]}(t, t')$ reducing to
\begin{equation}
\label{eq:GammaArb:SS}
\Gamma_{(-\infty, \tf]}(t, t')
	= \frac{1}{2 \ta^2 \lambda}
		\left[ \e^{-\lambda \lvert t - t' \rvert} - \frac{\kappa_-}{\kappa_+} \e^{-\lambda( 2\tf - t - t' )} \right]
\, ,
\end{equation}
the same expression as $\Gamma^\ind_{(-\infty, \tf]}(t, t')$ from eq.~\eqref{eq:GammaIndep:SS}.
This illustrates that the system loses its memory about the initial state as $\ti \to -\infty$.

Another comparison to our previous results from Sec.~\ref{sec:resultsPRX}
\cite{dabelow2019irreversibility} is obtained by
plugging the stationary state distribution
$p_\I(\VEC{\eta}_\I \given \VEC{x}_\I) = p_{\mathrm{ss}}(\VEC{\eta}_\I) = \sqrt{\ta/\pi} \, \e^{-\ta \VEC{\eta}_\I^2}$
into \eqref{eq:PathWeightArb:PosOnly} and performing the Gaussian integral over $\VEC{\eta}_\I$.
In that case, we should get back the result \eqref{eq:GammaInd}, \eqref{eq:PathWeightInd}
for independent initial conditions.
Indeed, including only the terms from \eqref{eq:PathWeightArb} which involve $\VEC{\eta}_\I$,
we evaluate the Gaussian integral over $\VEC{\eta}_\I$, yielding
\begin{align}
& \int \d\VEC{\eta}_\I \exp\left\{
	\int_\ti^\tf \!\! \d t
		\left[
			\frac{\sqrt{2\Da}}{2D}
			\left[ \dot{\VEC{x}}_t - \VEC{v}_t \right]^\T
			\frac{ \kappa_+ \e^{-\lambda (t - \ti)} - \kappa_- \e^{-\lambda(2\tf - t - \ti)} }
				 { \kappa_+ - \kappa_- \e^{-2 \lambda(\tf-\ti)} }
		\right] 
	\VEC{\eta}_\I
	\right.
\nonumber \\
& \qquad\qquad\qquad\qquad \left. \mbox{}	
	-  \frac{\ta}{2} \left[
			2 + \frac{  \frac{\Da}{D}(1 - \e^{-2\lambda(\tf-\ti)}) }{ \kappa_+ - \kappa_- \e^{-2 \lambda(\tf-\ti)} }
		\right] \VEC{\eta}_\I^2
	\right\}
\nonumber \\
& \qquad =
\exp\left\{ -\frac{1}{4D} \int_\ti^\tf \!\! \d t \int_\ti^\tf \!\! \d t' \;
		\left[ \dot{\VEC{x}}_t - \VEC{v}_t \right]^\T
		\left[ - \tfrac{\Da}{D} \Gamma_{[\ti,\tf]}^{\mathrm{ini}}(t, t') \right]
		\left[ \dot{\VEC{x}}_{t'} - \VEC{v}_{t'} \right]
	\right\}
\end{align}
with
\begin{equation}
\label{eq:PathWeightArb:GammaIni}
\Gamma_{[\ti,\tf]}^{\mathrm{ini}}(t, t')
= \left( \frac{1}{\ta} \right)
	\frac{ \kappa_+^2 \, \e^{-\lambda(t + t' - 2 \ti)} 
		 + \kappa_-^2 \, \e^{-\lambda (4 \tf - t - t' - 2 \ti)} 
		 - 2 \kappa_+ \kappa_- \, \e^{-2\lambda(\tf - \ti)} \left[ \e^{ \lambda(t - t')} + \e^{ \lambda(t' - t) } \right] }
	     { \left(\kappa_+ - \kappa_- \e^{-2 \lambda (\tf - \ti)} \right)
	       \left(\kappa_+^2 - \kappa_-^2 \e^{-2 \lambda (\tf - \ti)} \right) }
\, .
\end{equation}
A somewhat tedious but straightforward calculation then confirms
$\Gamma_{[\ti,\tf]}(t, t') + \Gamma_{[\ti,\tf]}^{\mathrm{ini}}(t, t') = \Gamma^\ind_{[\ti,\tf]}(t, t')$,
as expected.

\section{Irreversibility}
\label{sec:irr}
In stochastic thermodynamics \cite{Seifert:2008stp,Jarzynski:2011eai,Seifert:2012stf,van2015ensemble,seifert2018stochastic},
irreversibility is quantified by
comparing the probability $\pp{\Traj{\VEC{x}}} = \pp{\traj{\VEC x} | \VEC x_\I} p_\I(\VEC x_\I)$
of observing a
specific trajectory $\Traj{\VEC{x}} = \{\VEC{x}(t)\}_{t = \ti}^\tf$
in a given experimental setup
with the probability $\ppR{\TrajR{\VEC{x}}}$ of observing
the exact same trajectory traced out backwards when
providing identical experimental conditions.
In other words, $\ppR{\TrajR{\VEC{x}}}$ is the probability
of observing the ``time-reversed'' trajectory
\begin{equation}
\label{eq:xTilde}
\TrajR{\VEC{x}} = \{\tilde{\VEC{x}}(t)\}_{t = \ti}^\tf = \{\VEC{x}(\tf+\ti-t)\}_{t = \ti}^\tf
\, ,
\end{equation}
with $\tilde{\VEC{x}}(\ti)=\VEC{x}(\tf)$ and $\tilde{\VEC{x}}(\tf)=\VEC{x}(\ti)$,
under the time-reversed experimental protocol $\tilde{\VEC f}(\VEC x, t) := \VEC f(\VEC x, \tf + \ti - t)$
(note that we disregard for convenience the possibility that parts of the forces could be odd under time reversal;
it is straightforward to adapt the expressions below accordingly if necessary).
For passive Brownian motion, it has been shown that the log-ratio
of these path probabilities is related to the dissipation occurring along
the trajectory $\Traj{\VEC{x}}$, quantified as the total change of entropy
in the thermal bath and the system.
This fundamental connection makes the ``irreversibility measure''
\begin{equation}
\label{eq:DeltaSigma:def}
\Delta\Sigma[\Traj{\VEC{x}}] = -\kB \ln \frac{\ppR{\TrajR{\VEC{x}}}}{\pp{\Traj{\VEC{x}}}}
\end{equation}
a central quantity of interest also for active particles.
Indeed, its connection with dissipation and entropy is under
lively debate
\cite{Fodor:2016hff,marconi2017heat,mandal2017entropy,
Puglisi:2017crf,caprini2018comment,mandal2018mandal,dabelow2019irreversibility,caprini2019entropy}.

We here provide a general expression for $\Delta\Sigma$ based on our
result \eqref{eq:PathWeightArb}--\eqref{eq:PathWeightArb:PosOnly} for the path weight $\pp{\Traj{\VEC{x}}}$.
Since the time-reversed trajectory $\TrajR{\VEC{x}}$ is supposed to occur under identical
conditions as the forward trajectory $\Traj{\VEC{x}}$, we can express its probability
density via \eqref{eq:PathWeightArb}--\eqref{eq:PathWeightArb:PosOnly} as well,
if we replace $\VEC{v}(\VEC{x}, t)$ by the time-reversed protocol $\tilde{\VEC{v}}(\VEC{x}, t) = \tilde{\VEC{f}}(\VEC{x}, t)/\gamma$
(see below eq.~\eqref{eq:xTilde}).
Using \eqref{eq:xTilde} we then rewrite the path weight for the reversed 
path in terms of the forward path (and the original protocol $\VEC{v}_t=\VEC{v}(\VEC{x}(t),t)$).
The resulting expression for $\ppR{\TrajR{\VEC{x}}}$ is formally similar
to \eqref{eq:PathWeightArb},
just with the sign inverted for all $\dot{\VEC{x}}(t)$ terms and
all initial coordinates replaced by final ones.
Plugging the path weights $\pp{\Traj{\VEC{x}}}$ and $\ppR{\TrajR{\VEC{x}}}$
into \eqref{eq:DeltaSigma:def}, and denoting the
conditional average over the initial configuration $\VEC{\eta}_\I$ of the active fluctuations
\revised{$\int \d\VEC{\eta}_\I \, (\cdot) \, p_\I(\VEC{\eta}_\I \given \VEC{x}_\I)$} in
\eqref{eq:PathWeightArb:PosOnly} by $\langle \cdot \rangle_{\VEC{\eta}_\I \given \VEC{x}_\I}$
and the corresponding one over final configurations
\revised{$\int \d\VEC{\eta}_\F \, (\cdot) \, p_\F(\VEC{\eta}_\F \given \VEC{x}_\F)$}
by $\langle \cdot \rangle_{\VEC{\eta}_\F \given \VEC{x}_\F}$, we find
\begin{align}
\Delta\Sigma[\Traj{\VEC{x}}]
& = \frac{1}{T} \int_\ti^\tf \!\! \d t \int_\ti^\tf \!\! \d t' \;
		\dot{\VEC{x}}_t^\T \VEC{f}_{t'}
		\left[ \delta(t - t') - \tfrac{\Da}{D} \Gamma_{[\ti,\tf]}(t, t') \right]
	- \kB \ln \frac{p(\VEC{x}_\F)}{p(\VEC{x}_\I)}
\nonumber \\
& \qquad\qquad \mbox{}
	- \kB \ln
	\frac{ \left\langle \exp \left\{ -\int_\ti^\tf \d t
		\left[
			\frac{\sqrt{2\Da}}{2D}
			\left[ \dot{\VEC{x}}_t + \VEC{v}_t \right]^\T
			\frac{ \kappa_+ \e^{-\lambda (t - \ti)} - \kappa_- \e^{-\lambda(2\tf - t - \ti)} }
				 { \kappa_+ - \kappa_- \e^{-2 \lambda(\tf-\ti)} }
		    \VEC{\eta}_\F
		\right]
		\right\} \right\rangle_{\VEC{\eta}_\F \given \VEC{x}_\F} }
		{ \left\langle \exp \left\{ + \int_\ti^\tf \d t
		\left[
			\frac{\sqrt{2\Da}}{2D}
			\left[ \dot{\VEC{x}}_t - \VEC{v}_t \right]^\T
			\frac{ \kappa_+ \e^{-\lambda (t - \ti)} - \kappa_- \e^{-\lambda(2\tf - t - \ti)} }
				 { \kappa_+ - \kappa_- \e^{-2 \lambda(\tf-\ti)} }
		    \VEC{\eta}_\I
		\right]
		\right\} \right\rangle_{\VEC{\eta}_\I \given \VEC{x}_\I} }
\nonumber \\
& \qquad\qquad \mbox{}
	- \kB \ln \frac{ \left\langle \exp \left\{
		-\frac{\Da}{2D} \left[
			\frac{ \ta (1 - \e^{-2\lambda(\tf-\ti)}) }{ \kappa_+ - \kappa_- \e^{-2 \lambda(\tf-\ti)} }
		\right] \VEC{\eta}_\F^2
		\right\} \right\rangle_{\VEC{\eta}_\F \given \VEC{x}_\F} }
		{ \left\langle \exp \left\{
		-\frac{\Da}{2D} \left[
			\frac{ \ta (1 - \e^{-2\lambda(\tf-\ti)}) }{ \kappa_+ - \kappa_- \e^{-2 \lambda(\tf-\ti)} }
		\right] \VEC{\eta}_\I^2
		\right\} \right\rangle_{\VEC{\eta}_\I \given \VEC{x}_\I} }
\, .
\label{eq:DeltaSigma:res}
\end{align}
This expression constitutes the second central result of this work.
\revised{Given any spatial trajectory $\Traj{\VEC{x}} = \{\VEC{x}(t)\}_{t = \ti}^\tf$,
the measure $\Delta\Sigma[\Traj{\VEC{x}}]$ quantifies how irreversible this single
trajectory is in the sense of the definition \eqref{eq:DeltaSigma:def}.
A trajectory with $\Delta\Sigma[\Traj{\VEC{x}}]=0$ is reversible, i.e.\
movement of the AOUP forward or backward along the trajectory occurs with equal
probability, but the larger $\Delta\Sigma[\Traj{\VEC{x}}]$ the (exponentially) less likely it is
to observe the backward movement.}

Central properties of the active fluctuations which drive
the particle motion are represented by the parameters $\Da$ (the strength of the active
fluctuations) and $\ta$ (their correlation time, hidden in $\lambda=\sqrt{1+\Da/D}/\ta$).
Moreover, our general result \eqref{eq:DeltaSigma:res} contains averages over
the distributions of the active fluctuations $\VEC{\eta}_\I$ and $\VEC{\eta}_\F$
at the beginning of the particle
trajectory and at the beginning of the reversed trajectory
(see also \eqref{eq:PathWeightArb:PosOnly}).
We therefore presuppose that we have some knowledge or control
over these distributions when setting up the experiment, even though
%in typical experiments with active particles,
%measuring particle positions only,
the (microscopic) degrees of freedom related to the active fluctuations
typically are inaccessible, and so are specific realizations of $\VEC{\eta}(t)$
or the specific values of $\VEC{\eta}_\I$ and $\VEC{\eta}_\F$.
For artificial active colloids \cite{Bechinger:2016api},
or in computer experiments, we may imagine, e.g.,
to let the particles orient randomly before ``switching on''
the activity, possibly with a specific strength (distribution).

In the spirit of quantifying irreversibility by asking how likely
it is to observe a reversed trajectory compared to its forward twin
when starting from \emph{identically prepared experimental setups} (except for
the initial particle position, which is $\VEC{x}_\I$
for the forward path and $\VEC{x}_\F$ for the backward path),
we may take the distributions for
$\VEC{\eta}_\I$ and $\VEC{\eta}_\F$
to be the same, or to be ``mirror images'' of each other under sign-inversion,
depending on the physical situation modeled by the active fluctuations $\VEC{\eta}(t)$
(see the discussion in \cite{dabelow2019irreversibility})
\footnote{In fact, to us these seem to be the only proper choices,
if we want $\Delta\Sigma$ to quantify \emph{irreversibility}.
For arbitrary, unrelated distributions of $\VEC{\eta}_\I$ and $\VEC{\eta}_\F$,
we would compare forward and backward paths generated under
different experimental conditions.}.
Moreover, we may imagine the experiment to be prepared in a way
that the initial distributions of
the active fluctuations for forward and backward motion are independent of particle positions
(a notable exception arising, if the experiment starts from a joint steady state).
For such independent initial conditions with identical (or ``mirrored'') distributions,
the third line in \eqref{eq:DeltaSigma:res} vanishes.
The second line, however, is still non-zero, and can
be interpreted to quantify the contribution to irreversibility from
the initial configuration of the active fluctuations.

The first line in \eqref{eq:DeltaSigma:res} is independent of $\VEC{\eta}_\I$ and $\VEC{\eta}_\F$,
and thus measures the irreversibility associated with the time-evolution of the spatial
particle position alone.
It contains three terms (two in the double-integral and a boundary term),
which all represent different contributions to irreversibility.
The boundary term $-\kB \ln [{p(\VEC{x}_\F)}/{p(\VEC{x}_\I)}]$
does not involve any parameters characterizing the thermal bath
or the active fluctuations, and
is usually interpreted as the change in system entropy of the AOUP
between the beginning and end of the
trajectory $\Traj{\VEC{x}}$ \cite{dabelow2019irreversibility}.
The integral involving $\delta(t-t')$
is independent of the active parameters
$\Da$ and $\ta$, and is formally identical to the entropy
produced in the thermal bath along the trajectory $\Traj{\VEC{x}}$
as known for \emph{passive} Brownian motion \cite{Seifert:2012stf}.
However, in the present case of an AOUP it does not capture the full heat
dissipation, because in addition to the force $\VEC{f}(\VEC{x}(t),t)$ also active
self-propulsion ``forces'' $\sqrt{2\Da}\VEC{\eta}(t)$ drive the particle and
contribute to dissipation, i.e.\
even though the $\delta(t-t')$-integral can be interpreted as the
``thermal contribution'' to irreversibility due to the AOUP being in contact with a heat bath,
it cannot be identified with the entropy produced in this thermal environment
(see also the detailed discussion in \cite{dabelow2019irreversibility}).
The second, proper double-integral
encodes the (statistical) characteristics of the active fluctuations
via its kernel $\Gamma_{[\ti,\tf]}(t, t')$ and, furthermore, vanishes if
active propulsion is ``switched off'', i.e.\ for $\Da=0$. Hence, it can be
interpreted to measure the irreversibility ``produced'' by the active fluctuations
along the trajectory $\Traj{\VEC{x}}$, and we will
refer to it as the ``active contribution'' to irreversibility.

These two contributions to irreversibility from the particle trajectory
$\Traj{\VEC{x}}$, the thermal one and the active one,
are non-zero only if external forces $\VEC{f}(\VEC{x},t)=\gamma \VEC{v}(\VEC{x},t)$
are present \emph{in addition to the active self-propulsion}
(likewise for the integral in the second line, i.e.\ for the contribution associated
with the initial preparation of the system),
implying that the trajectories of ``free active motion'' appear reversible.
For non-conservative forces, both contributions
typically lead to a time-extensive increase of irreversibility with trajectory length $\tau$.
For conservative forces $\VEC{f}(\VEC{x})=-\VEC{\nabla}U(\VEC{x})$ derived from
a stationary confining potential $U(\VEC{x})$, the thermal contribution reduces to the boundary term
$[U(\VEC{x}_\I)-U(\VEC{x}_\F)]/T$ and thus is non-extensive in $\tau$.
Due to the double-integral nature of the active part with the non-local kernel
$\Gamma_{[\ti,\tf]}(t, t')$ a similarly obvious argument does not apply.
Indeed, the question whether or not, or in how far, the trajectory of an AOUP in a confining potential
appears (ir)reversible is still not fully answered
\cite{Fodor:2016hff,dabelow2019irreversibility,dabelow2020preprint}.
We can, however, draw some conclusions from considering the limiting cases
of small and large correlations times $\ta \to 0$ and $\ta \to \infty$.
In the first case, $\ta \to 0$, the active fluctuations become white and thus behave like
a thermal bath, such that the AOUP can be imagined to be a passive
Brownian particle in contact with
a heat bath at effective temperature $\gamma (D+\Da)/\kB$,
trapped in a confining potential.
Accordingly, irreversibility production is non-extensive.
In the second case, $\ta \to \infty$, the active fluctuations become constant
and thus behave like
a bias force which slightly tilts the confining potential.
Again, the situation is similar to a trapped passive Brownian particle with
non-extensive irreversibility production.
We can therefore expect that the active contribution to irreversibility
in a confining potential
may become maximal at some intermediate value of $\ta$.

Another important implication of the result~\eqref{eq:DeltaSigma:res}
is that the \emph{rate} at which irreversibility is produced in the stationary state
(i.e., upon letting $\ti \to -\infty$, c.f.\ Section~\ref{sec:PathWeightIndependentSS})
is independent of the specific initial distribution $p_{\mathrm i}(\VEC x_{\mathrm i}, \VEC\eta_{\mathrm i})$.
Indeed, the terms in the second and third lines of~\eqref{eq:DeltaSigma:res} vanish as $\ti \to -\infty$,
and the memory kernel $\Gamma_{[\ti,\tf]}(t,t')$ in the first line assumes the form~\eqref{eq:GammaArb:SS},
independent of the initial distribution (see the discussion in Section~\ref{sec:p0etaArbitrary}).
Hence, if we are only interested in long-time properties,
the initial configuration in particular of the self-propulsion drive is irrelevant.
While we might have intuitively expected 
that the long-time irreversibility production rate is independent of
the details of the initial setup,
it is not completely obvious in the presence of memory effects
\cite{harris2009current,puglisi2009irreversible}.
The fact that we can verify it for AOUPs is reassuring though,
in so far as control over the initial state is limited in typical active particle systems
(as already mentioned above).
For infinitely long trajectories $\tf \to \infty$
in the stationary state $\ti \to -\infty$, the expression
for $\Delta\Sigma[\Traj{\VEC{x}}]$ reduces further to
\begin{equation}
\Delta\Sigma_{(-\infty,\infty)}[\Traj{\VEC{x}}]
= \frac{1}{T} \int_\ti^\tf \!\! \d t \int_\ti^\tf \!\! \d t' \;
		\dot{\VEC{x}}_t^\T \VEC{f}_{t'}
		\left[ \delta(t - t') - \tfrac{\Da}{D} \Gamma_{(-\infty,\infty)}(t, t') \right]
%	- \kB \ln \frac{p(\VEC{x}_\F)}{p(\VEC{x}_\I)}
\end{equation}
with $\Gamma_{(-\infty,\infty)}(t, t')$ from \eqref{eq:GammaIndep:SSinf}
(see the discussion around eq.~\eqref{eq:GammaArb:SS}),
in agreement with the findings in \cite{caprini2019entropy,zamponi2005fluctuation}.

\revised{We conclude this discussion with a remark concerning the
relation between the expression for $\Delta\Sigma[\Traj{\VEC{x}}]$ found earlier
in \cite{dabelow2019irreversibility} and the result~\eqref{eq:DeltaSigma:res}
derived here.
In \cite{dabelow2019irreversibility}, we have calculated
$\Delta\Sigma[\Traj{\VEC{x}}]$ under the assumption
that the active fluctuations are in their stationary state,
independent of initial particle positions,
at the beginning of the forward path and at the beginning of the backward path,}
i.e., $p_\I(\VEC{\eta}_\I \given \VEC{x}_\I) = p_{\mathrm{ss}}(\VEC{\eta}_\I) = \sqrt{\ta/\pi} \, \e^{-\ta \VEC{\eta}_\I^2}$
\revised{and
$p_\F(\VEC{\eta}_\F \given \VEC{x}_\F) = p_{\mathrm{ss}}(\VEC{\eta}_\F) = \sqrt{\ta/\pi} \, \e^{-\ta \VEC{\eta}_\F^2}$.}
The resulting expression for \eqref{eq:DeltaSigma:res}
looks formally identical to the first line in \eqref{eq:DeltaSigma:res}, but with
$\Gamma_{[\ti,\tf]}(t, t')$ substituted by $\Gamma^\ind_{[\ti, \tf]}(t, t')$
from \eqref{eq:GammaInd}
\revised{(compare with eqs.~(42b) and (40a) in \cite{dabelow2019irreversibility}).}
As we can see from the calculation at the end of Section \ref{sec:p0etaArbitrary}
above, the ``amount of irreversibility'' stemming from the particular stationary-state initial
configuration of the active fluctuations has been absorbed into $\Gamma^\ind_{[\ti, \tf]}(t, t')$,
and is therefore not explicitly visible in \cite{dabelow2019irreversibility}
as an additional term analogous to the second line in \eqref{eq:DeltaSigma:res}.

\section{Conclusions}
\label{sec:discussion}
What can we learn about the non-equilibrium nature of an active system
by observing particle trajectories, i.e., the
evolution of particle positions over time?
Within the framework of a minimal model for particulate  active matter
on the micro- and nanoscale,
the active Ornstein-Uhlenbeck particle
\cite{Fily:2012aps,Farage:2015eii,Maggi:2014gee,Argun:2016nbs,maggi2017memory,chaki2018entropy,Marconi:2015tas,shankar2018hidden}
(see eqs.~\eqref{eq:EOM}, \eqref{eq:LangevinActiveNoise}), we
here contribute an essential step towards exploring
this question by deriving an exact analytical expression
for the path weight (eqs.~\eqref{eq:PathWeightArb}, \eqref{eq:GammaArb}, \eqref{eq:PathWeightArb:PosOnly}),
which is valid for any values of the
model parameters, any external driving forces,
arbitrary initial particle positions and configurations
of the active fluctuations, and arbitrary trajectory durations.
We use this general expression to calculate the log-ratio of path weights
for forward versus backward trajectories (see eq.~\eqref{eq:DeltaSigma:res}).
In analogy to the stochastic thermodynamics of passive Brownian particles
\cite{Seifert:2008stp,Jarzynski:2011eai,Seifert:2012stf,van2015ensemble,seifert2018stochastic},
such an irreversibility measure may provide an approach
towards a thermodynamic description of active matter
\cite{Fodor:2016hff,marconi2017heat,mandal2017entropy,
Puglisi:2017crf,caprini2018comment,mandal2018mandal,dabelow2019irreversibility,caprini2019entropy}.

In future works we may build on these results to further
explore the non-equilibrium properties of AOUPs.
A highly interesting problem is a possible thermodynamic interpretation
of the path probability ratio $\Delta\Sigma[\Traj{\VEC{x}}]$
\cite{dabelow2019irreversibility}, e.g.,
via exploring its connection to active pressure \cite{takatori2014swim,solon2015pressure},
to the different phases observed in active matter \cite{Cates:2015mip},
or to the arrow of time
\cite{roldan2015decision,roldan2018arrow} in these systems.
Such a thermodynamic interpretation, in particular concerning the role of dissipation,
may finally allow to quantify efficiency fluctuations in stochastic heat engines
operating between active baths \cite{Krishnamurthy:2016ams},
in analogy to passive stochastic heat engines
\cite{Verley:2014tuc,Verley:2014uto,manikandan2019efficiency}.
Other important questions which can be approached directly by
using our general result for the path weight $\pp{\Traj{\VEC{x}}}$
include the analysis of the response behavior under external perturbations
\cite{dal2019linear}
or of violations of the fluctuation-response relation
\cite{harada2005equality,gnesotto2018broken} due to the
inherent non-equilibrium character of active matter,
and their potential for probing properties of the active
fluctuations \cite{gnesotto2018broken}.
\revised{
Finally, it would be interesting to explore if our analytical methods
used here to integrate out the simple Ornstein-Uhlenbeck fluctuations \eqref{eq:LangevinActiveNoise}
can be extended to treat more general active fluctuations, like the ones
considered in \cite{sevilla2019generalized}.}

\begin{acknowledgments}
We thank Stefano Bo for stimulating discussion.
L.D.\ gratefully acknowledges support by the Nordita visiting PhD program.
This research has been funded by the
Swedish Research Council (Vetenskapsr{\aa}det) under the Grants No. 2016-05412 (R.E.)
and by the Deutsche Forschungsgemeinschaft (DFG) within the Research Unit FOR 2692 under Grant No.~397303734 (L.D.).
\end{acknowledgments}
%%%%%%%%%%%%

\appendix

\section{Evaluation of $\Gamma_\sigma(t,t')$}
We here outline the calculation of $\Gamma_\sigma(t,t')$ as the inverse
of the differential operator
\begin{equation}
\label{eq:app:PathWeightGauss:V}
V_\sigma(t, t')
	:= \delta(t - t') \left[
		- \ta^2 \partial_{t'}^2 + 1 + \frac{\Da}{D}
		+ \delta(t') \left( -\ta^2 \partial_{t'} - \ta + \frac{1}{\sigma^2} \right)
		+ \delta(\tau - t') \left(\ta^2 \partial_{t'} + \ta \right)
	\right]
\, ,
\end{equation}
i.e., $\Gamma_\sigma(t, t')$ is a solution of the equation
$\int_0^\tau \d t' \; V_\sigma(t, t') \Gamma_\sigma(t', t'') = \delta(t - t'')$.
In fact, the operator $V_\sigma(t, t')$ is ``diagonal'' in the time arguments such that
$\Gamma_\sigma(t,t')$ solves the differential equation
\begin{equation}
\label{eq:app:DGLGamma}
\left[
	- \ta^2 \partial_{t}^2 + 1 + \frac{\Da}{D}
	+ \delta(t) \left( -\ta^2 \partial_{t} - \ta + \frac{1}{\sigma^2} \right)
	+ \delta(\tau - t) \left(\ta^2 \partial_{t} + \ta \right)
\right] \Gamma_\sigma(t, t') = \delta(t - t')
\, .
\end{equation}
Note that $t'$ is essentially a fixed parameter here, just like $D$, $\Da$, $\ta$ and $\sigma$.
To find the solution, we follow the procedure from \cite{dabelow2019irreversibility}, i.e.,
we compose $\Gamma_\sigma(t, t')$ from two parts as $\Gamma_\sigma(t, t') = G(t, t') + H_\sigma(t, t')$.
First, we construct the function $G(t,t')$ as the Green's function solving
the equation
$[-\tau_a^2 \partial_t^2 + (1 + \Da/D) ] G(t, t') = \delta(t-t')$
with homogeneous boundary conditions $G(0, t') = G(\tau, t') = 0$.
Second, we determine
$H_\sigma(t,t')$ as a solution
of the homogeneous problem
$[-\tau_a^2 \partial_t^2 + (1 + \Da/D)] H_\sigma(t, t') = 0$
such that the boundary terms are fixed as prescribed by \eqref{eq:app:DGLGamma}.

We can construct both parts, $G(t, t')$ and $H_\sigma(t, t')$, from the general solution
\begin{equation}
\label{eq:app:Gamma}
\Gamma(t) = a^+ \e^{\lambda t} + a^- \e^{-\lambda t}
\, , \quad
\lambda = \frac{1}{\ta} \sqrt{1+\frac{\Da}{D}}
\,, \quad
a^\pm  = \mathrm{const}
\end{equation}
of the homogeneous ordinary differential equation
\begin{equation}
\left[ -\ta^2 \partial_{t}^2 + 1 + \frac{\Da}{D} \right]\Gamma(t) = 0
\end{equation}
associated with \eqref{eq:app:DGLGamma}.
The Green's function $G(t,t')$ is exactly the same as in \cite{dabelow2019irreversibility}.
Accordingly, its construction is completely analogous to the procedure outlined
in Appendix B of \cite{dabelow2019irreversibility},
and we only recall the result here,
\begin{equation}
\label{eq:app:G}
G(t,t') = \frac{1}{2\ta^2\lambda}
	\frac{\e^{\lambda(\tau-|t-t'|)} - \e^{\lambda(\tau-t-t')} + \e^{-\lambda(\tau-|t-t'|)} - \e^{-\lambda(\tau-t-t')}}
		 {\e^{\lambda\tau} - \e^{-\lambda\tau}}
\, .
\end{equation}
The difference between the present calculation and the one in \cite{dabelow2019irreversibility}
is the boundary term at $t=0$, which
contains a contribution from the arbitrary (Gaussian)
initial distribution of the active fluctuations.
To take both boundary terms in~\eqref{eq:app:DGLGamma} into account,
we make an ansatz of the form \eqref{eq:app:Gamma}
for the function $H_\sigma(t, t')$, i.e.,
$H_\sigma(t,t') = a^+ \e^{\lambda t} + a^- \e^{-\lambda t}$. The coefficients $a^\pm$ are fixed
by ensuring
that the full solution $\Gamma_\sigma(t,t') = G(t,t')+H_\sigma(t,t')$ fulfills
\eqref{eq:app:DGLGamma}. Plugging $G(t,t')+H_\sigma(t,t')$ into \eqref{eq:app:DGLGamma},
and using $[-\tau_a^2 \partial_t^2 + (1 + \Da/D) ] G(t, t') = \delta(t-t')$
and $\left[ -\ta^2 \partial_{t}^2 + 1 + \frac{\Da}{D} \right]H_\sigma(t,t') = 0$,
we are left with
\begin{multline}
	\delta(t) \left[ -\ta^2 \partial_{t} G(t,t')|_{t=0} + a^+(1/\sigma^2-\ta\kappa_+) + a^-(1/\sigma^2-\ta\kappa_-) \right]
\\ \mbox{}
  + \delta(\tau - t) \left[ \ta^2 \partial_{t} G(t,t')|_{t=\tau} + a^+\ta\kappa_+\e^{\lambda\tau} + a^-\ta\kappa_-\e^{-\lambda\tau} \right]
= 0
\, ,
\end{multline}
where $\kappa_\pm = 1 \pm \lambda\ta = 1 \pm \sqrt{1+\Da/D}$.
Requiring that the terms in the two square brackets each vanish, we can solve for the coefficients $a^\pm$,
yielding
\begin{eqnarray}
a^+ & = & \left( \frac{1}{\ta} \right) \frac{ (1-\sigma^2 \ta \kappa_-) \left[ \e^{-\lambda( 2\tau - t' ) } -  \e^{ -\lambda (2 \tau + t') } \right] - \kappa_- \left[ \e^{-\lambda( 2\tau + t' )} - \e^{-\lambda(4\tau - t') } \right] }{ \kappa_+ (1-\sigma^2 \ta \kappa_-) - \kappa_- (1-\sigma^2 \ta \kappa_+) \, \e^{-2 \lambda \tau} } \,, \\
a^- & = & \left( \frac{1}{\ta} \right) \frac{ -(1 - \sigma^2 \ta \kappa_+) \left[ \e^{-\lambda( 2\tau - t' ) } - \e^{ -\lambda (2 \tau + t') } \right] + \kappa_+ \left[ \e^{-\lambda t'} - \e^{-\lambda(2\tau - t') } \right] }{ \kappa_+ (1-\sigma^2 \ta \kappa_-) - \kappa_- (1-\sigma^2 \ta \kappa_+) \, \e^{-2 \lambda \tau} } \,.
\end{eqnarray}
Substituting these coefficients
into the above ansatz for
$H_\sigma(t,t')$ [see below~\eqref{eq:app:G}] and
combining it with $G(t, t')$ from~\eqref{eq:app:G} according to
$\Gamma_\sigma(t,t') = G(t,t')+H_\sigma(t,t')$,
we obtain the result stated in eq.~\eqref{eq:Gamma:Sigma} of the main text.

\bibliography{msPreprint}

\end{document}